\newcommand{\be}{\begin{equation}}
\newcommand{\ee}{\end{equation}}
\newcommand{\bea}{\begin{eqnarray}}
\newcommand{\eea}{\end{eqnarray}}

\documentclass[prb,showpacs]{revtex4}
\usepackage{amssymb}
\usepackage{graphicx}
\usepackage{dcolumn}
\usepackage{bm}
\usepackage{amsmath}
\usepackage{amsfonts}

\begin{document}

\title{Self-consistent approach for the quantum confined Stark effect in shallow quantum wells}
\author{I. V. Ponomarev}
\email{ilya@physics.qc.edu}
\author{L. I. Deych}
\author{A. A. Lisyansky}

 \affiliation{Department of Physics, Queens College of CUNY,
Flushing, NY 11367}
\date{\today }

\begin{abstract}
A computationally efficient, self-consistent complex scaling
approach to calculating characteristics of excitons in an external
electric field in quantum wells is introduced. The method allows
one to extract the exciton resonance position as well as the
field-induced broadening for the exciton resonance. For the case
of strong confinement the trial function is represented in
factorized form. The corresponding coupled self-consistent
equations, which include the effective complex potentials, are
obtained. The method is applied to the shallow quantum well. It is
shown that in this case the real part of the effective exciton
potential is insensitive to changes of external electric field up
to the ionization threshold, while the imaginary part has
non-analytical field dependence and small for moderate electric
fields. This allows one to express the exciton quasi-energy at
some field through the renormalized expression for the zero-field
bound state.
\end{abstract}
 \maketitle

%\preprint{APS/123-QED}

%%%%%%%%%%%%%%%%%%%%%%%%%%%%%%%%%%%%%%%%%%%%%%%%%%%%%%%%%%%%%%%%%%%%%%%
\section{Introduction\label{sec:sec1}}
The quantum-confined Stark effect (QCSE) has attracted a lot of
attention since  its discovery\cite{Miller84,Miller85a} due to its
diverse optoelectronics applications (see for example
Ref.\onlinecite{Miller90}). In quantum well (QW) structures with a
layer thickness less than the effective exciton Bohr radius, an
electron and a hole are forced to orbit closer to each other, and
the binding energy of the electron-hole pair increases by a factor
between two and three. This additional stability makes exciton
resonances in QW's observable even at room temperatures. More
important, these resonances are still observable in a uniform
electric field perpendicular to the growth direction, which is as
large as 50 times the classical ionization field for a
three-dimensional exciton. The latter also produces a relatively
large red shift in the absorption peak up to $2.5$ times the
binding energy of the exciton, while the width of the exciton
resonance grows only moderately.
%Another important consequence of the quantum confined
% is that these resonances survive in a constant electric
%field, which can be as large as 50 times of classical ionization
%field for three-dimensional exciton, if this filed is applied
%perpendicular to the growth direction of the structure.
%As a result, exciton energy can be shifted by the electric field
%by up to $2.5$ times of the binding energy of the exciton, while
%the width of the exciton resonance grows only moderately.
Such a behavior is opposite to the properties of 3-d excitons or
QW excitons in a field parallel to the plane of the
QW.\cite{Miller84,Miller85a} The origin of this phenomenon stems
from the fact that the electric field induced decay time for QW
excitons is  determined not by its binding energy ($\sim 10$ meV
in GaAs/AlGaAs quantum wells), but rather by the quantum well
barriers' heights ($\sim 300-400$ meV).

Despite the fact that the QCSE was discovered almost 20 years ago
and qualitatively is well understood, there is no complete
quantitative solution for the problem so far. Usually, it is
treated numerically with different degrees of approximation. The
conventional approach consists in  the separation of the problem
into two consecutive steps. The first step in modelling QCSE is to
omit the Coulomb interaction of an electron-hole pair and to find
single particle resonance positions for an electron and a hole
separately. In moderate fields (up to $10^5$ V/cm) and non-shallow
quantum wells (quantum wells with more than one discrete level
inside) the lowest-lying resonances are so sharp that they are
often approximately treated as bound states, neglecting the
possibility of tunnelling.
% In this case electrons and holes can be
%thought of as remaining in bound states (at least for lowest
%energy states) despite of the presence of the electric field.
These simple methods include either an infinite well
approximation,\cite{Bastard83a,Miller85a,Sanders87a,Ahn87a,Miller85a}
where the effect of penetration of the electron (hole) wave
functions under the barrier is taken into account by introducing
an effective QW width,\cite{Ahn87a,Miller85a} or variational
calculations for finite potential wells\cite{Bastard83a,Brum85},
which produce only the real part of the exciton quasi-energy.
Within a weak-field regime, the Stark shift is a small
second-order effect in the perturbation theory sense. The absolute
value of a shift decreases with a decrease of the QW width.
Achieving large shifts requires strong fields, where the
perturbation theory fails. In this regime a significant broadening
of the states appears due to the field induced tunnelling.
Therefore, it is necessary to devise non-perturbative methods for
the correct determination of the exciton resonance energies and
tunnelling life-times in these systems in the presence of a strong
field.

There are several ways of approaching this problem mathematically.
The single-particle quasi-bound states in QW's in the presence of
a uniform electric field have been studied with the help of the
time-independent Schr\"{o}dinger equation using Airy functions
with real and complex arguments\cite{Ahn86a,Kuo99a,Gonzalez96a},
stabilization technique\cite{Borondo86a,Porto94a,Panda99a}, finite
difference method\cite{Harrison99a} and complex coordinate
method\cite{Bylicki96a}. Another group of methods utilizes a
dynamical approach\cite{Panda99a,Juang90a,Panda96a} where the time
evolution of a tunnelling wave packet is simulated by numerical
integration of the time-dependent Schr\"{o}dinger equation.

It is worth  pointing out that all of the above mentioned
calculations of decaying quasi-bound states are made in the
single-particle approximation without consideration of the Coulomb
attraction between the electron and the hole. Excitonic effects
are incorporated at the next step of the calculations, which
usually take into account only a real part of the exciton binding
energy shift, and are performed with the help of variational
methods\cite{Miller85a,Brum85,Sanders87a,Nojima88a}.

The goal of this paper is twofold. First, we  develop an
alternative method of studying QCSE that allows us to calculate
both the position of the exciton resonances and their
field-induced tunnelling life-times  on equal footing. This method
is a generalization of a self-consistent approach  developed
recently by the authors\cite{pon03a} for the exciton in QW without
the electric field. Second, we introduce the model of QCSE in
shallow quantum wells (SQW), which allows us to obtain analytical
results for the resonance positions and tunnelling widths as well
as for the modified exciton "ground state" wave function. In this
model we approximate SQW by a delta-functional potential.
Calculations are significantly simplified in this model due to the
fact that the single-particle one-dimensional problem for a
particle in a delta-potential and in an applied uniform electric
field can be solved exactly.

The paper is organized as follows. In Sec. \ref{sec:2} we
generalize the self-consistent approach of Ref.\onlinecite{pon03a}
for the presence of the electric field, and derive main
self-consistent field equations for complex resonance energy
positions for QCSE, using the method of complex scaling. In Sec.
\ref{sec:3} we apply our approach to the model of SQW  for
calculations of the effective exciton potential. One of the
results of these calculations is the phase diagram representing
regions of exciton's stability in strong electric fields for
different well's widths and/or barrier potential.  Concluding
remarks are given in Sec. \ref{sec:4}.

\section{Complex scaling self-consistent approach for QCSE\label{sec:2}}
One of the goals of this paper is to develop a new method for
calculating excitonic characteristics in QW in the presence of the
electric field due to QCSE. In order to present the method clearly
we consider the simplest model of QW excitons. In particular, we
assume that both conduction and valence bands are non-degenerate,
and that they both have an isotropic parabolic dispersion
characterized by the masses $m_e$ and $m_{h}$, respectively. Thus,
we neglect valence-band mixing\cite{Ekenberg87}, anisotropy of the
system, and dielectric\cite{Gerlach98a} and effective mass
mismatch between the well and barrier materials. All these
effects, which are very important for realistic computations can
be, however, easily incorporated into our approach. Throughout the
paper we use effective atomic units (a.u.), which means that all
distances are measured in the units of the effective Bohr radius
$a_B=\hbar^2\epsilon/\mu^*e^2$, energies in the units of
$\mu^*e^4/\hbar^2\epsilon^2\equiv 2\textrm{ Ry}$, and masses in
the units of reduced electron-hole mass $\mu^*$, where
$1/\mu^*=1/m_e^*+1/m_h^*$, and $\epsilon$ is an average background
dielectric constant of the well and barrier materials. In these
notations $m_{e,h}=m_{e,h}^*/\mu^*$, where $m_{e,h}^*$ are
effective masses of an electron and a heavy hole.

Within the approximations explained above, and after the standard
procedure of excluding the center-of-mass of the perpendicular
motion in the plane of the layers\cite{Miller85a}, the excitonic
Hamiltonian is given by
\begin{eqnarray}
\hat{H} &=&E_g+H_e+H_h+K_r+V_{reh},\label{H1}\\
H_e(z_e) &=& -\frac{1}{2m_e}\frac{\partial^2}{\partial z_e^2}
+U_1(z_e)-Fz_e,\nonumber\\
H_h(z_h) &=& -\frac{1}{2m_h}\frac{\partial^2}{\partial z_h^2}
+U_2(z_h)+Fz_h,\nonumber\\
K_r(r) &=& -\frac{1}{2}\left[\frac{\partial^2}{\partial
r^2}+\frac{1}{r}\frac{\partial}{\partial r}\right],\nonumber\\
V_{reh}(R) &=& -\frac{1}{\sqrt{r^2+(z_e-z_h)^2}}\equiv
-\frac{1}{R} \nonumber,
\end{eqnarray}
where $E_g$ is a gap energy, $z$ is the growth direction, $r$
measures a relative electron-hole distance in the transverse
direction $r=\sqrt{(x_e-x_h)^2+(y_e-y_h)^2}$, $U_{1,2}$ are the
quantum well confining potential in $z$ direction for the electron
and the hole, respectively, and $F=|e|\textit{E}$ measures the
strength of the electric field. We also assume that the
quasi-bound exciton state is independent of the angle in the $xy$
plane, and exclude the corresponding term from the kinetic energy
of the relative motion $K_r$.

Strictly speaking, because of the possibility of tunnelling, there
are no bound states in a system with uniform electric field, and
the energy spectrum of Hamiltonian (\ref{H1}) is continuous: $E\in
\ (-\infty,\infty)$. This means that the respective wave functions
do not vanish at infinity and, therefore, are not normalizable  in
$L^2$ sense. Then a proper approach to the problem is to consider
a scattering problem. The scattering matrix $S$ contains all the
physical information about the system. For example, resonances
correspond to the poles of the $S$-matrix in the lower half of the
complex energy plane. It is well known that these complex energies
can also be  found by solving an eigenvalue problem for the
time-independent Schr\"{o}dinger equation, when the wave functions
are chosen to have only outgoing components, but do not have an
incident wave (so called Siegert or Gamow wave functions).
Formally, after matching the usual "bound-state" boundary
conditions one can obtain the correct complex eigenvalues.
However, the eigenfunctions corresponding to these complex
eigenvalues  exponentially diverge. This circumstance prohibits an
application of a standard variational principle to calculations of
approximate values of the quasi-bound states for QCSE.

The complex coordinate (complex scaling) approach (see for review
Refs. \onlinecite{Reinhart82a,Bardsley78a}) is devised to
circumvent this problem. The idea consists in applying the
coordinate transformation $\mathbf{R}\rightarrow
\mathbf{R}\exp(i\gamma)$ to all coordinates in the Hamiltonian
(\ref{H1}). Such a transformation makes the new Hamiltonian
non-Hermitian [$\hat{H}\rightarrow \hat{H}(\gamma)$] and changes
its spectrum respectively. The continuous spectrum
moves\cite{Herbst78} to $-i\infty$. In addition, a discrete set of
complex points lying below the real energy appears. These complex
quasi-energies correspond to the tunnelling resonances under
consideration. It is important that the eigenfunctions
$\Psi_{\gamma}(\mathbf{R})$ corresponding to these complex
eigenvalues $W=E-i\Gamma/2$ are square-integrable, and, therefore,
can be calculated by means of the variational principle, which
requires the stationarity of the functional
\begin{equation}\label{funW}
    W\left[\Psi\right]=\int\;\Psi_{\gamma}\hat{H}(\gamma)\Psi_{\gamma}\,dV\left/
    \int\;\Psi_{\gamma}\Psi_{\gamma}\,dV.\right.
\end{equation}
It should be noted, however, that the biorthogonality
normalization (without complex conjugation of the left
eigenvector) is used in the definition of the energy functional
instead of the standard scalar product. The complex quasi-energies
should also be stationary with respect to changes in rotation
angle $\gamma$. The method of complex scaling and methods similar
to it are widely and successfully used in studies of atomic and
molecular auto-decaying states\cite{Reinhart82a,Bardsley78a}.

 The new variational principle, Eq. (\ref{funW}), for the rotated non-Hermitian
Hamiltonian can be used for extension of the self-consistent
method, recently developed by the authors\cite{pon03a}, to the
problem of QCSE. According to this method, we construct an
approximate exciton wave function $\Psi_{\gamma}(z_e,z_h,r)$ with
the help of the unknown functions $\psi_1, \psi_2,\ldots$, where
each function $\psi_k$ depends on a fewer number of variables than
the total wave function. Considering variations of these functions
independently, from the variational principle, Eq.~(\ref{funW}),
we obtain coupled integro-differential equations for $\psi_k$. The
resulting coupled equations can be solved by successive
approximations. The convergence of this procedure allows for
estimating to what degree a given functional dependence of the
trial function is close to the exact wave function. In the case of
zero electric field and sufficiently strong localization (the
exciton "$z$-size" is smaller than its three-dimensional Bohr
radius), we found\cite{pon03a} that representation of the trial
function in the form of a product of functions of different
coordinates gives an excellent approximation for exciton energies
even after only one iteration. We expect, therefore, that a
similar form will give accurate results for quasi-bound states in
the presence of the electric field, at least, when it is not very
large. Thus,  we choose the trial function for the Hamiltonian
(\ref{H1}) in the form of the product:
\begin{equation}\label{MFtrialF}
  \Psi_{\textrm{trial}}(r,z_e,z_h)=\psi(r)\chi_e(z_e)\chi_h(z_h).
\end{equation}
Assuming normalization of every function in this product, we
substitute function Eq.~(\ref{MFtrialF}) in Eq.~(\ref{funW}), vary
each function in a product separately, and obtain the system of
coupled integro-differential equations
\begin{eqnarray}
\left[e^{-2i\gamma}K_r(r)+e^{-i\gamma}\overline{V}_r(r)\right]\psi(r)&=&W_X\psi(r),\label{mfeq1}\\
\left[H_e(e^{i\gamma}z_e)+e^{-i\gamma}\overline{V}_e(z_e)\right]\chi_e(z_e)&=&W_e\chi_e(z_e),\label{mfeq2}\\
\left[H_h(e^{i\gamma}z_h)+e^{-i\gamma}\overline{V}_h(z_h)\right]\chi_h(z_h)&=&W_h\chi_h(z_h),\label{mfeq3}
\end{eqnarray}
where the following notations for effective potentials are
introduced:
\begin{eqnarray}
\overline{V}_r(r)&=&\langle \chi_e\chi_h|-1/R|\chi_e\chi_h \rangle,\label{efpotr}\\
\overline{V}_{e,h}(z_{e,h})&=&\langle
\psi\chi_{h,e}|-1/R|\psi\chi_{h,e} \rangle. \label{efpotz}
\end{eqnarray}
The angle brackets of biorthogonal inner product imply that the
integration of the Coulomb potential with corresponding wave
functions is carried out over two of three independent variables.

 Solving the system of equations (\ref{mfeq1})--(\ref{mfeq3}) we obtain the best
approximation for the entire wave function in the form of a
product (\ref{MFtrialF}). The corresponding value of the total
quasi-bound energy $W=E-i\Gamma/2$ is given by Eq. (\ref{funW}),
which can be rewritten in the form
\begin{equation}\label{totalE}
W=\langle \Psi|\hat{H}|\Psi\rangle=W_e+W_h+W_X-
\langle\chi_e|e^{-i\gamma}\overline{V}_{e}|\chi_e\rangle-\langle\chi_h|e^{-i\gamma}\overline{V}_{h}|\chi_h\rangle.
\end{equation}

Eigenfunctions and corresponding complex eigenvalues of Eqs.
(\ref{mfeq1})--(\ref{mfeq3}) can be found with the help of
successive iterations. As in the case of zero electric field, the
zero order solution is obtained after neglecting the effective
Coulomb terms [$\overline{V}^{(0)}_{e,h}$=0] in single-particle
Eqs. (\ref{mfeq2}) and (\ref{mfeq3}):
\begin{equation}
H_{e,h}(e^{i\gamma}z_{e,h})\chi_{e,h}^{(0)}(z)=W_{e,h}^{(0)}\chi_{e,h}^{(0)}(z).\label{mfeq20}
\end{equation}
With the help of functions $\chi_{e,h}^{(0)}$ a zero order
approximation for the effective potential
$\overline{V}^{(0)}_{r}(r)$ is calculated and substituted into Eq.
(\ref{mfeq1}). This effective potential has the meaning of a
two-dimensional effective electron-hole interaction induced by
collective effects of an attractive Coulomb interaction, quantum
well potentials in the $z$ direction, and an applied uniform
electric field.
 The first quasi-bound state for this effective two-dimensional exciton is found from the following equation
 \begin{equation}\label{mfeq10}
  \left[e^{-2i\gamma}K_r+e^{-i\gamma}\overline{V}^{(0)}_r(r)\right]\psi^{(0)}(r)=W^{(0)}_X\psi^{(0)}(r).\\
\end{equation}
The eigenfunction $\psi^{(0)}(r)$ is then  substituted into Eqs.
(\ref{efpotz})  to calculate  a new approximation
$\overline{V}^{(1)}_{e,h}(z_{e,h})$ for the effective potentials.
This process can be continued until the potentials are
self-consistent to a high order of accuracy, i.e. until the
condition
\begin{equation}
\langle\psi|\overline{V}_{r}|\psi\rangle\approx
\langle\chi_e|\overline{V}_{e}|\chi_e\rangle\approx\langle\chi_h|\overline{V}_{h}|\chi_h\rangle
\label{sscond}
\end{equation}
is fulfilled.

The main advantage of the approach described above  is that it
allows one to calculate, in principle, not only field-induced
single-particle widths $\Gamma_{e,h}$ but the \emph{exciton width}
$\Gamma_X$ as well, which describes a renormalization of the
electron-hole pair life-time by the effective interaction. For
moderate fields this width is a non-analytical function of $F$,
and  is small compared to $\Gamma_{e,h}$. The exact calculation of
this contribution is possible only numerically. The details of
such calculations for particular semiconductor structures will be
presented elsewhere. Below we consider the properties of the
solution of Eqs.~(\ref{mfeq1})--(\ref{mfeq3}) obtained after the
first iteration of the self-consistent method for the case of a
shallow quantum well, where some analytical results can be
derived.

It should be stressed that the most important feature of the
complex scaling method is the biorthogonality normalization. The
other feature, the coordinate rotation itself
$\mathbf{R}\rightarrow \mathbf{R}\exp(i\gamma)$, is necessary to
guarantee the convergence of the integrals
(\ref{efpotr},\ref{efpotz}). Although, this rotation is important
to negate the asymptotic divergence of the \emph{exact}
eigenfunctions of the quasi-bound state, it was
found\cite{Bardsley78a} that in the case of sharp resonances (when
the width $\Gamma$ is small in comparison with the real part of
the energy) the optimal angle of rotation $\gamma$ is close to
zero. Moreover, the final answer for quasi-energy is not very
sensitive to the change of $\gamma$. Therefore, for approximate
calculations we will fix $\gamma=0$ and choose the trial functions
to be square integrable from the very beginning. However, we will
permit variational parameters to be \emph{complex}. The latter
assumption allows us to extract the resonance width in the
problem, while still keeping the problem tractable.  We will
follow this procedure in the subsequent sections.
%%%%%%%%%%%%%%%%%%
\section{$\delta$-functional model for a shallow quantum well in an electric field\label{sec:3}}
 A zero approximation of a self-consistent field approach
for the electron (hole) part of the wave function in the $z$
direction, Eqs.~(\ref{mfeq20}), is equivalent to the standard
problem of a single-particle quasi-bound state in QW. Therefore,
its solution can be obtained with the help of any method described
previously\cite{Ahn86a,Kuo99a,Gonzalez96a,Borondo86a,Porto94a,Panda99a,Harrison99a,Bylicki96a,Juang90a,Panda96a}.
Let us first solve the problem exactly using Airy functions with
complex arguments.

We define a shallow quantum well as a well in which only one bound
state exists for both electrons and holes. It is well-known that,
if a well's width, $L$, tends to zero, a shallow quantum well can
be successfully described by a $\delta$-functional well
\begin{equation}\label{deltapotential}
  V(z)=U-\alpha\delta(z),
\end{equation}
where $\alpha=UL$ is characterizes the $\delta$-potential strength
and $U$ is the barrier height. For convenience, the first constant
term $U$ on the left hand side of Eq. (\ref{deltapotential}) will
be omitted hereafter. In terms of the total Hamiltonian, Eq.
(\ref{H1}), it means that the energy band gap, $E_g$, is the
barrier's energy band gap:
$E_g^{\textrm{bar}}=E_g^{\textrm{well}}+U_e+U_h$. The
delta-functional approximation is directly applicable only either
to very narrow QW\cite{pon03a,Andreani97a} or to wells with a
small band-gap offset,\cite{Andreani97a} i.e. when the well width
and/or the band offsets are very small so that the carrier wave
functions are mostly in the barrier region. However, the range of
its applicability can be extended\cite{pon03a} up to the moment of
the appearance of the second bound state by introducing the
effective strength of the potential
\begin{equation}\label{alphaeffective}
  \alpha=UL_{\textrm{eff}}=\sqrt{\frac{2U}{m}-\frac{\pi^2x_1^2}{m^2L^2}}
\end{equation}
where $x_1$ is the solution of the equation
\begin{equation}
  x=1-2/\pi\arcsin\left(\frac{\pi x}{\sqrt{2mUL^2}}\right).\label{transeq}
\end{equation}
$L_{\textrm{eff}}$ is chosen in such a way that the energy of the
bound state in the $\delta$-potential matches the ground state
energy of the finite well problem\footnote{Another way to define
the effective potential strength $\alpha$ is to apply a
variational method to the finite quantum well problem with a trial
function $\chi_{trial}(z)=\exp(-\kappa z)$.}. In some sense, the
model of the $\delta$-functional QW is complementary to the model
of an effective infinite quantum well\cite{Miller85a} (EIQW),
which is used to approximate finite QW with large widths (or
barrier heights) when the number of levels in a well is large.
Indeed, the more discrete levels exist in the QW the better the
EIQW model works for the ground state, but it fails
%gives a wrong eigenstate dependence on $L$,
when the well has only one discrete level. On the other hand, the
$\delta$-functional QW is not applicable for quantum wells with
more than one level. For typical parameters in
Al$_{0.3}$Ga$_{0.7}$As/GaAs structures the range of applicability
of $\delta$-functional SQW extends up to a well width $L\approx
40\AA$.

\begin{figure}[tbp]
\includegraphics{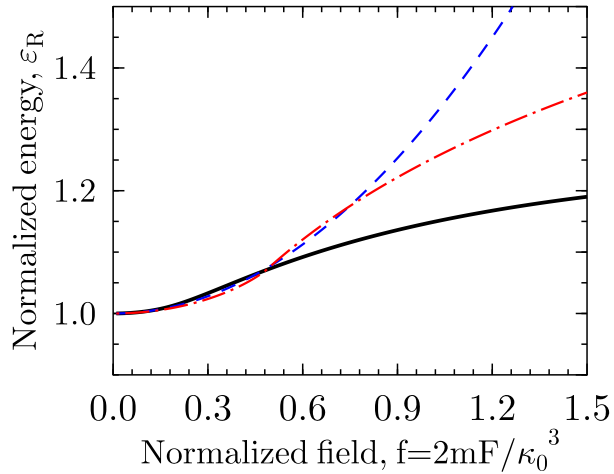}
\caption{The real part $\varepsilon_R$ of the normalized energy
$\kappa^2$ as a function of normalized electric field
$f=2mF/\kappa_0^3$. The solid line represents the exact solution
of Eq. (\ref{airy_eneq}). The dashed line shows the quadratic
behavior of an approximate solution, Eq. (\ref{airy_en_sm}) for
small $f$. The dot-dashed line is the real part for the energy,
Eq. (\ref{airy_en_var}), obtained with the help of a variational
method with a trial function (\ref{airy_trial_f}). See below in
text.} \label{fig1}
\end{figure}

\begin{figure}[tbp]
\includegraphics{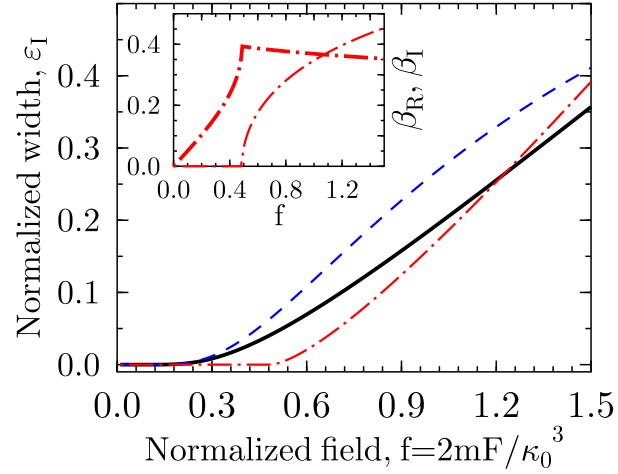}
\caption{The imaginary part $\varepsilon_I$ of the normalized
energy $\kappa^2$ as a function of normalized electric field
$f=2mF/\kappa_0^3$. The solid line represents the exact solution
of  Eq. (\ref{airy_eneq}). The dashed line shows the exponential
non-analytical growth of an approximate solution, Eq.
(\ref{airy_en_sm}).  The dot-dashed line is the imaginary part for
the energy, Eq. (\ref{airy_en_var}), obtained with the help of a
variational method with a trial function (\ref{airy_trial_f}). The
insert shows the behavior of the real and the imaginary parts of
the variational parameter $\beta$} for the trial function
(\ref{airy_trial_f}). \label{fig2}
\end{figure}

The advantages of the $\delta$-functional SQW model become
especially obvious in the case when  a uniform electric field is
applied. The simplicity of the model allows one to derive
analytically \emph{both} the electric field induced energy shift,
and the tunnelling life-time. Let us first find the complex energy
and the corresponding wave function of the single one-particle
quasi-stationary state for the electron (the case of a hole is
obtained by substituting $-F$ for $F$). The respective
one-dimensional Schr\"{o}dinger equation is
\begin{equation}\label{airy01}
\left[-\frac{1}{2m}\frac{d^2}{d\, z^2}
-\alpha\delta(z)-Fz\right]\chi(z)=E\,\chi(z).
\end{equation}
It is convenient to introduce the dimensionless variables for the
field and the wave vector,
\begin{equation}\label{dimensionless_f}
f=2mF/\kappa_0^3,  \qquad \kappa=\sqrt{-2mE}/\kappa_0,
\end{equation}
where $\kappa_0=m\alpha$ is the ground state wave vector for an
unperturbed $\delta$-potential problem. After performing the
standard coordinate transformation
\begin{equation}\label{y2airy02}
  y=\kappa_0f^{1/3}\left(z-\frac{\kappa^2}{\kappa_0 f}\right)
\end{equation}
we obtain
\begin{equation}\label{airy03}
\frac{d^2\,\chi(y)}{d\,z^2}+y\chi(y)=
-2f^{-1/3}\delta(y+y_0)\chi(y),
\end{equation}
where $y_0=\kappa^2 f^{-2/3}$. Everywhere except the point $z=0$
(or equivalently, $y=-y_0$),  Eq. (\ref{airy03}) coincides with
the Airy equation.\cite{Abramowitz} In order to find the
quasi-bound state we have to use boundary conditions at infinity,
which require that the solution does not contain incoming waves.
It means that we construct our solution in the form of an
evanescent wave to the left, and propagating outgoing wave to the
right:
\begin{equation}
  \chi =\left\{
\begin{array}{lccc}
C_1(\kappa)\left[\textrm{Ai}(-y)-i\textrm{Bi}(-y) \right], &
y\rightarrow +\infty &\displaystyle{\frac{C_1}{\sqrt{\pi}y^{1/4}}
\exp\left(i2/3y^{3/2}\right)}& z>0\\
C_2(\kappa)\textrm{Ai}(-y),& y\rightarrow -\infty &
\displaystyle{\frac{C_2}{\sqrt{\pi}|y|^{1/4}}
\exp\left(-2/3|y|^{3/2}\right)}
 & z<0\label{AiryL}
\end{array}
\right.
 \end{equation}
Note that propagating asymptotic for the right part of the wave
function is valid  for $z \gg \kappa^2/\kappa_0 f$, i.e. far away
from the origin $z=0$. Near the origin, the wave function
decreases exponentially for both $z<0$ and $z>0$. For small fields
the evanescent asymptotic can be used to describe the behavior of
the wave function at both sides of the point $z=0$. The unknown
constants $C_{1,2}$ and complex energy are found after matching
the boundary conditions at $y=-y_0$. For the normalized energy
$\kappa^2$ we obtain the following equation:
\begin{equation}\label{airy_eneq}
\textrm{Ai}\left(\frac{\kappa^2}{f^{2/3}}\right)\textrm{Bi}\left(\frac{\kappa^2}{f^{2/3}}\right)
+i\textrm{Ai}^2\left(\frac{\kappa^2}{f^{2/3}}\right)=\frac{f^{1/3}}{
2\pi}.
\end{equation}
It has one complex solution
\begin{equation}\label{airy_en_gen}
  \varepsilon_c\equiv\kappa^2=\varepsilon_R(f)+i\varepsilon_I(f).
\end{equation}
The real and the imaginary parts numerically calculated from
Eq.~(\ref{airy_eneq}) are shown in Figs.~\ref{fig1},~\ref{fig2}.
The notion of the quasi-bound state has sense, when the width
$\varepsilon_I$ is small in comparison with the real part of the
energy. This happens at small fields, i.e. when the condition
$f^{1/3} < 1$ is fulfilled. In this case the right hand side of
the Eq. (\ref{airy_eneq}) is small, and the real part of the
argument of the Airy functions is large: $\textrm{Re}(\kappa^2
f^{-2/3})\gg 1$. With the help of asymptotic expansions of the
Airy functions,\cite{Abramowitz} Eq. (\ref{airy_eneq}) can be
reduced to
\begin{equation}\label{airy_eqsm}
  \kappa=1+\frac{5}{32}\frac{f^2}{\kappa^6}+\frac{1155}{2048}\frac{f^4}{\kappa^{12}}+\cdots+
  \frac{i}{2}\exp\left(-\frac{4\kappa}{3f}\right)\left(1-\frac{5f}{24\kappa}+\cdots\right).
\end{equation}
Solving this equation by iterations, we obtain after the first
successive approximation
\begin{equation}\label{airy_en_sm}
  \varepsilon_R\approx 1+\frac{5}{16}f^2,\quad \varepsilon_I\approx
  \exp\left(-\frac{4}{3f}\right).
\end{equation}
For comparison we present these results in
Figs.~\ref{fig1},~\ref{fig2} along with the results of the exact
numerical solution of Eq.(\ref{airy_eneq}). The first two terms in
the expansion of the wave function (\ref{AiryL}) near the origin
with respect to the field parameter $f$ are
\begin{equation}
\chi(z)\approx \sqrt{\kappa_0}e^{-\kappa_0 |z|}\left[1+\frac{f}{4}
\textrm{sign}(z)|\kappa_0 z|(1+|\kappa_0 z|)\right].
\end{equation}
%%%%%%%%%%%%%%%%%%%%%%%%%%%%%%%%%%%%%%%%%%%%%%%%%%%%%%%%%%%%%%%%%%%
%%%%%%%%%%%%%%%%%%%%%%%%%%%%%%%%%%%%%%%%%%%%%%
%%%%%%%%%%%%%%%%%%%%%%%%%%%%%%%%%%%%%%%%%%%%%%
The result for the real part of the energy presented in Eq.
(\ref{airy_en_sm}) can  also be obtained from the second order of
the standard quantum mechanical perturbation theory:
\begin{equation}\label{airy_stark_per}
  E_0^{(2)}=-2m\int_0^{\infty}\frac{|\langle
\chi_{k,odd}^{(0)}(z)|-Fz|\chi_{0}^{(0)}(z)
\rangle|^2}{\kappa_0^{2}+k^2}dk,\\
\end{equation}
where $\chi_{k,odd}^{(0)}(z)=\sin(kz)/\sqrt{\pi}$ are the odd wave
functions of the unperturbed Hamiltonian. The non-analytical
exponential dependence of the field induced broadening in
Eq.(\ref{airy_en_sm}) is typical for the Stark effect in a
hydrogen-like atom $\Gamma\propto\exp\left[-(4/3)(
E_0/F\ell)\right]$, where $E_0$ and $\ell$ are the absolute value
of energy and the average radius of localization for the ground
state without the field. In the case of QW excitons, the main
contribution to $E_0$ comes from the single particle confinement
energy, which is much larger than the exciton ground state energy
$E_0/E_X\sim U^2L^2/\textrm{Ry}^2a_B^2\gg 1$. Because of this
circumstance, QCSE in QW is observable for very large fields.

\begin{figure}[tbp]
\includegraphics{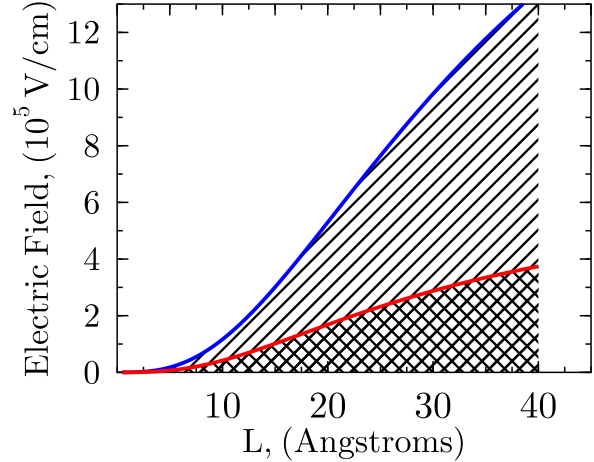}
\caption{The phase diagram $(L,F)$ for the region, where the
quasi-bound states for an electron (upper curve) and a hole (lower
curve) exist. The data are presented for
GaAs-Ga$_{0.7}$Al$_{0.3}$As SQW. The calculations are based on the
following physical constants: $m_e^*=0.067m_0$, $m_h^*=0.45m_0$,
$U_e=340$ meV, $U_h=70$ meV, $\epsilon=13.8$. The upper border for
widths, $L=40\AA$, is a limit of applicability of
$\delta$-functional model.} \label{fig3}
\end{figure}

In a strong enough electric field, the broadening quickly grows
and becomes comparable to the real part of energy. In this case
the resonance associated with the quasi-bound state disappears,
becoming invisible on the background of the continuous  spectrum.
The condition of the existence of the resonance can be
approximately formulated as $f\leq 1$.
%\begin{equation}\label{cond_disap}
%  f\leq 1.
%\end{equation}
Since $f=2mF/(mUL_{eff})^3$, it is easy to determine the region of
parameters where quasi-bound states (and QCSE respectively) are
observable. In Fig.~\ref{fig3} we present the phase diagram for
the region of the existence of the quasi-bound states as a
function of well width, $L$, and electric field for a
GaAs-Ga$_{0.7}$Al$_{0.3}$As SQW. As expected, the larger fields
and smaller widths destroy the localization of the exciton in the
quantum well.
%%%%%%%%%%%%%%%%%%%%%%%%%%%%%%%%%%%%%%%%%%%%%%%%%%%%%%%%%%%%%%%%%%%
%%%%%%%%%%%%%%%%%%%%%%%%%%%%%%%%%%%%%%%%%%%%%%

A knowledge of the complex energy and the corresponding wave
function, Eq. (\ref{AiryL}), allows us, in principle, to find the
corrections to the quasi-bound energy due to the effective
electron-hole interaction. To this end, one would have to perform
a coordinate rotation $z_{e,h}\rightarrow z_{e,h}\exp(i\gamma)$ to
provide the convergence at infinity for the electron and the hole
eigenfunctions (\ref{AiryL}), to  substitute these wave functions
into the integral in Eq. (\ref{efpotr}), and calculate the
effective complex potential for the electron-hole interaction.

 Unfortunately, the exact forms for the wave functions
 (\ref{AiryL}) lead to fairly cumbersome numerical calculations of the integral
(\ref{efpotr}). However, the simple and rather accurate analytical
estimates can be obtained with the help of the variational wave
function of the following form
\begin{equation}\label{airy_trial_f}
\chi_{trial}(z;\beta)=\sqrt{\kappa_0(1-\beta^2)}\exp\left[\kappa_0\left(-|z|+\beta
z\right)\right].
\end{equation}
A function of similar form has been successfully used in the past
to model the real part correction to the single particle energy in
quantum wells with finite width and in the presence of the
electric field\cite{Brum85}. In our approach, however, we do not
restrict the variational parameter $\beta$ to be a real number,
extending the range of its values on the whole complex plane. The
only restriction is $\textrm{Re}(\beta)<1$, which is dictated by
the square integrability of the biorthogonal scalar product.
Applying the variational principle to the Eq. (\ref{airy01}) we
obtain the following equations for the energy
\begin{equation}\label{airy_var_en}
  \kappa^2(\beta)=1-\beta^2+\frac{f\beta}{1-\beta^2}.
\end{equation}
The corresponding equation for the variational parameter $\beta$
is
\begin{equation}\label{airy_en_var}
  -f/2+\beta-f/2\beta^2-2\beta^3+\beta^5=0.
\end{equation}
For small fields, the last three terms on the left hand side of
Eq. (\ref{airy_en_var}) can be omitted, and we obtain
$\beta\approx f/2$. The corresponding value for energy shift
$\varepsilon_R-1\approx f^2/4$, which differs from Eq.
(\ref{airy_en_sm}) only by a factor of $1.25$. For moderate values
of the electric field this variational energy is even closer to
the exact value than the perturbative result, Eq.
(\ref{airy_en_sm}). It is more important, however, that Eq.
(\ref{airy_en_var}) also allows one to analyze the field induced
decay of the electron. In order to understand how the electron
acquires a finite life-time, one needs to include the third
quadratic term in Eq. (\ref{airy_en_var}). The solution of the
modified equation takes the form
\begin{equation}\label{airy_en_var_root2}
  \beta=f^{-1}\left[1-\sqrt{1-f^2}\right],
\end{equation}
from which it is seen that when the electric field exceeds a
critical value $f=f_{cr}=1$, a square-root singularity appears,
and the corresponding energy acquires an imaginary part. The exact
solution of Eq. (\ref{airy_en_var}) changes the type of singular
behavior of function $\beta(f)$ in the vicinity of $f_{cr}$, and
lowers the critical field value
\begin{equation}\label{fcr}
  f_{cr}=0.487.
\end{equation}
 The dependences of the real and
imaginary parts of variational energy, Eq (\ref{airy_var_en}), on
the normalized electric field, obtained from the solution of Eq.
(\ref{airy_en_var}) are shown in Figs.~\ref{fig1} and~\ref{fig2}.
We can see that the variational results for the real part of the
energy are in very good agreement (especially in the case of small
and moderate fields) with the results of the exact solution of Eq.
(\ref{airy_eneq}). The variational method with a simple trial
function (\ref{airy_trial_f}) also captures the effect of the
field-induced decay of excitons characterized by a non-analytical
dependence of the decay rate upon the electric field.
%The quantitative comparison of the results for the
%imaginary part, however, is not so impressive. The actual fast
%growth of the decay rate begins at the lower fields, and, of
%course, it has different, exponential type of singularity, that
%cannot be approximated by a polynomial of any finite degree.
Thus, the inclusion of the complex solutions of Eq.
(\ref{airy_en_var}) for the variational parameter gives
substantial physical insight into the problem of calculation of
the effective exciton potential. We have to stress here that in
the  above procedure we performed only partial optimization of the
functional (\ref{funW}), since we fixed $\gamma=0$. The
variational optimization of  Eq. (\ref{mfeq20}) with respect to
the rotation angle would give a non-zero optimal angle
$\gamma_{opt}$ and even better agreement with the exact solution
of  Eq. (\ref{airy_eneq}). As we already mentioned, in the case of
sharp resonances\cite{Bardsley78a} the optimal angle of rotation
$\gamma$ is small. Moreover, the final answer for quasi-energy is
not very sensitive to the change of $\gamma$. Therefore, the full
optimization gives minimal corrections to the obtained result for
the complex energy and variational parameter $\beta$, making,
however, the further calculations of the exciton effective
potential very cumbersome.

\section{Effective potential and binding energy corrections for exciton\label{sec:4}}
The effective potential for a quasi 2D exciton is given by the
integral
\begin{equation}\label{Veff01}
V_{\textrm{eff}}(r;\kappa_0^{(e)},\kappa_0^{(h)},F)=-\int_{-\infty}^{\infty}dz_e
\int_{-\infty}^{\infty}dz_h  \frac{\left|\chi_e(z_e)\right|^2
\left|\chi_h(z_h)\right|^2}{\sqrt{r^2+(z_e-z_h)^2}}.
\end{equation}
 For electron and hole wave functions we will use
functions in a form Eq. (\ref{airy_trial_f}) with corresponding
values of the complex variational parameter $\beta$.
 For further calculations it is useful to introduce the following
 notations:
we will denote wave numbers, variational parameters, and wave
functions, related to the electron (hole) by index $1(2)$, for
instance wave numbers $\kappa_i\equiv\kappa_{0}^{(e,h)}$, where
$i=1$ or $2$. We also introduce renormalized parameters
$\kappa_{iL,R}$ according to
\begin{eqnarray}
\kappa_{1L,R}&=&\kappa_1\left(1\pm \beta_1\right)\label{kappa1LR}\\
\kappa_{2L,R}&=&\kappa_2\left(1\mp \beta_2\right)\label{kappa2LR}\\
\beta_i&=&\beta_i(2m_iF/\kappa_i^3)=\beta_{iR}+i\beta_{iI},\quad \beta_R,\beta_I\geq 0.   \\
\end{eqnarray}
With these notations the wave functions (\ref{airy_trial_f}) and
effective potential, Eq. (\ref{efpotr}), can be rewritten as
\begin{equation}
  \chi_i(z_i) =\sqrt{\frac{\kappa_{iR}\kappa_{iL}}{\kappa_{i}}}\left\{
\begin{array}{cc}
\exp(\kappa_{i\,L}z), & z<0\\
\exp(-\kappa_{i\,R}z), & z>0\label{Airy_var_f02}
\end{array}
\right.
 \end{equation}
\begin{eqnarray}\label{Veff01F2}
\overline{V}^{(0)}_r(r;\kappa_1,\kappa_2,F)&=&
-\int_{0}^{\infty}dz_1\int_{0}^{\infty}dz_2\left[
\frac{\chi_{1L}^2\chi_{2L}^2+\chi_{1R}^2\chi_{2R}^2}{\sqrt{r^2+(z_1-z_2)^2}}+
\frac{\chi_{1L}^2\chi_{2R}^2+\chi_{1R}^2\chi_{2L}^2}{\sqrt{r^2+(z_1+z_2)^2}}
\right]\nonumber\\
  &\equiv &
 V_{I}(\kappa_{1L},\kappa_{2L},r)+V_{I}(\kappa_{1R},\kappa_{2R},r)+
  V_{II}(\kappa_{1L},\kappa_{2R},r)+V_{II}(\kappa_{1R},\kappa_{2L},r),
\end{eqnarray}
where the last line introduces notations for each of four terms
comprising the integral representation for $\overline{V}^{(0)}_r$.
As we had shown before\cite{pon03a} the potentials $V_{I,II}$ can
be represented with the help of the function
%through the function The
%functions $V_{I,II}$ have the same form as in Ref.\cite{pon03a},
%where the following representation for the respective integrals
%was derived
\begin{equation}\label{TE}
T_{\alpha}\equiv T(\kappa_{\alpha} r)=\int_{0}^{\infty}
\frac{\exp(-2\kappa_{\alpha}
t)dt}{\sqrt{r^2+t^2}}=\frac{\pi}{2}\left[\textbf{H}_0(2\kappa_{\alpha}
r)-\textrm{Y}_0(2\kappa_{\alpha} r) \right],
\end{equation}
where $\textbf{H}_0$ is the zero-order Struve function and
$\textrm{Y}_0$ is the zero-order Neumann or Bessel function of the
second kind.\cite{Abramowitz} Then
\begin{equation}\label{V1V2}
V_{I,II}(\kappa_{\alpha},\kappa_{\beta})=-\frac{\kappa_{1L}\kappa_{1R}\kappa_{2L}\kappa_{2R}}{2\kappa_1\kappa_2}
 \left[\frac{T(\kappa_{\beta} r)\pm T(\kappa_{\alpha} r)}{\kappa_{\alpha}\pm\kappa_{\beta}}\right],
\end{equation}
and the final result for the effective exciton potential induced
by QW confinement and the presence of the electric field takes the
form
\begin{equation}\label{VeffFr}
\overline{V}^{(0)}_r(r;\kappa_1,\kappa_2,F)=\frac{\kappa_1\kappa_2(1-\beta_1^2)(1-\beta_2^2)}{2}\left[
\frac{T_{1L}+T_{2L}}{\kappa_{1L}+\kappa_{2L}}+\frac{T_{1R}+T_{2R}}{\kappa_{1R}+\kappa_{2R}}+
\frac{T_{1L}-T_{2R}}{\kappa_{2R}-\kappa_{1L}}+\frac{T_{1R}-T_{2L}}{\kappa_{2L}-\kappa_{1R}}\right].
\end{equation}
 Without the electric field Eq.(\ref{VeffFr}) is reduced to the
 result obtained in Ref.\cite{pon03a}
\begin{equation}\label{Vr0}
\overline{V}^{(0)}_r(r;\kappa_1,\kappa_2,0)=-\frac{2\kappa_1\kappa_2}{\kappa_2^2-\kappa_1^2}
\left[\kappa_2 T(\kappa_1 r)- \kappa_1 T(\kappa_2 r)\right].
\end{equation}
where it was shown that the properties of this potential are
governed by a single parameter
\begin{equation}\label{dkappa}
  d^2=\frac{1}{2}\left(\frac{1}{\kappa_1^2}+\frac{1}{\kappa_2^2}\right)
  =\langle \chi_1(z_1)\chi_2(z_2)|(z_1-z_2)^2|\chi_1(z_1)\chi_1(z_2)\rangle.
\end{equation}
This parameter has a simple physical meaning of the average square
of the distance between the electron and the hole in the $z$
direction. It controls the transition between small $r$ and large
$r$ asymptotics of the zero-field potential (\ref{Vr0}):
\begin{equation}
  \overline{V}^{(0)}_r(r;\kappa_1,\kappa_2,0) \thickapprox \left\{
\begin{array}{cc}
\frac{1}{d}\ln(r/d), & r\lesssim d,\\
-\frac{1}{\sqrt{r^2+d^2}},  & r\gtrsim d.\label{Vr0_asympt}
\end{array}
\right.
 \end{equation}
The effective zero-field electron-hole interaction for the exciton
in the SQW starts from the true logarithmic Coulomb potential of a
point charge in two dimensions that smoothly transforms at
distances $r\sim d$ to the screening potential (\ref{Vr0_asympt})
with three-dimensional Coulomb tails.

The understanding of the role played by the parameter $d$ helps to
substantially simplify calculations of the ground state energy.
For example, we found that the variational energy obtained with a
single parameter trial function
\begin{equation}\label{trpsi_scheq}
\varphi_{trial}=
\frac{2\exp(d/\lambda)}{\sqrt{\lambda(\lambda+2d)}}
\exp(-\sqrt{r^2+d^2}/\lambda)
\end{equation}
coincides with  more sophisticated numerical approaches with
excellent accuracy. The respective expression for the ground state
energy can be written as a sum of two terms:
$E_X=E_X^{(0)}(\lambda)+E_X^{(1)}$, where
\begin{equation}\label{excenergy_sc0}
E_X^{(0)}(\lambda)=-\frac{2}{\lambda}\frac{1}{1+2d/\lambda}+\frac{1}{\lambda^2}
\left[1-\frac{(2d/\lambda)^2E_1(1,2d/\lambda)\exp(2d/\lambda)}{1+2d/\lambda}\right],
\end{equation}
with $\textrm{E}_1(x)$ being  the exponential
integral.\cite{Abramowitz}, and
\begin{equation}\label{excenergy_sc1a}
  E_X^{(1)} \approx
-\frac{2d}{\lambda^2}\left[\frac{1}{2}-0.557\frac{2d}{\lambda}+0.563
\left(\frac{2d}{\lambda}\right)^2+\cdots\right].
\end{equation}
The first term $E_X^{(0)}$ should be taken for the optimal value
of $\lambda$ that brings Eq.~(\ref{excenergy_sc0}) to the minimum.
The second term, Eq. (\ref{excenergy_sc1a}), is a small
correction, which is calculated for a given optimal $\lambda.$

\begin{figure}[tbp]
\includegraphics{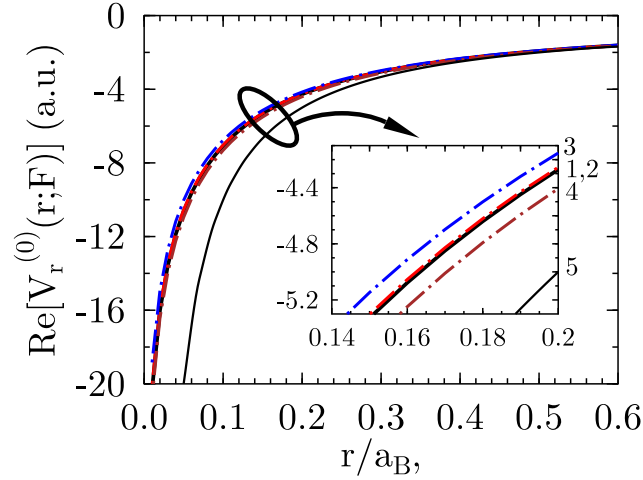}
\caption{The profile of the real part of the effective
self-consistent exciton potential
$\overline{V}^{(0)}_r(r;\kappa_1,\kappa_2,F)$ for different values
of the electric field (dot-dashed lines). The thick solid line
(curve 5) below the other curves represents $-1/r$ behavior, and
is given for comparison. The insert is a magnified view of the
circled area. The labels for the curves correspond to, curve 1,
$F=0$ V/cm, curve 2, $F=0.4\times 10^{5}$ V/cm, curve 3,
$F=0.8\times 10^{5}$ V/cm, curve 4, $F=4\times 10^{5}$ V/cm. All
data are for $d/a_B=0.16$ that corresponds $L=20\AA$ finite
quantum well in AlGaAs/GaAs materials.} \label{fig4}
\end{figure}

\begin{figure}[tbp]
\includegraphics{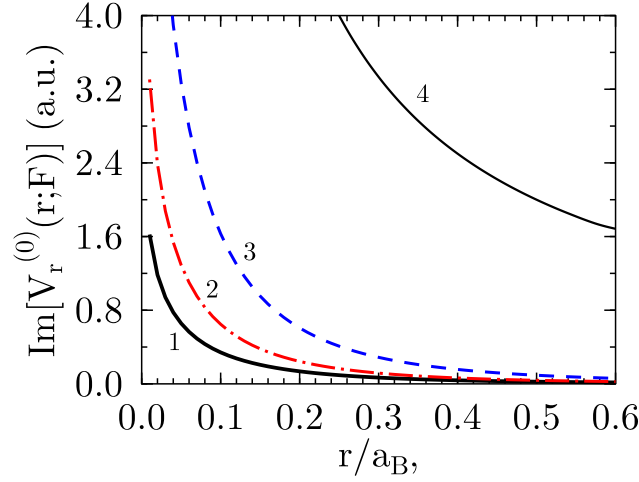}
\caption{The profile of the imaginary part of the effective
self-consistent exciton potential $\overline{V}^{(0)}_r(r;F)$  for
different values of electric field (dot-dashed lines). The upper
solid line (curve 4) represents $1/r$ behavior, and is given for
comparison. The labels for the curves correspond to, curve 1,
$F=10^5$ V/cm, curve 2, $F=2.\times 10^{5}$ V/cm, curve 3,
$F=4\times 10^{5}$ V/cm. All data are for $d/a_B=0.16$ that
corresponds to $L=20\AA$ finite quantum well in AlGaAs/GaAs
materials.} \label{fig5}
\end{figure}

In order to understand how these results are modified by the
electric field we study the properties of the effective potential
 Eq.(\ref{VeffFr}). Figures~\ref{fig4} and \ref{fig5} show the profiles of the real and the imaginary parts of
 this potential calculated for several
different electric fields. For concreteness we chose for
calculations parameters typical for GaAs-Ga$_{0.7}$Al$_{0.3}$As
SQW with width $L=20\AA$. In our calculations we used the
following physical constants: $m_e^*=0.067m_0$, $m_h^*=0.45m_0$,
$U_e=340$ meV, $U_h=70$ meV, $\epsilon=13.8$. For these constants
we have $\kappa_1=5.68$, $\kappa_2=7.31$, and $d=0.16$. The latter
value tells us that in a zero field the exciton is strongly
localized in the QW. The critical fields $F^{cr}_{1,2}$ derived
from Eqs. (\ref{fcr},\ref{dimensionless_f}) are
$F^{cr}_{h}=0.82\times 10^5$V/cm and $F^{cr}_{e}=2.57\times
10^5$V/cm. According to our phase diagram, Fig.~\ref{fig3}, a
field that larger than $1.7\times 10^5$V/cm should be considered
as strong, since in this case the hole  has already a substantial
probability to tunnel through the barrier, escaping an interaction
with the electron.
%%%%%%%%%%%%%%%%%%%%%%%%%%%%%%%%%%%%%%%%%%%%%%%%%%%%%%%%%%%%%%%%%%%
%%%%%%%%%%%%%%%%%%%%%%%%%%%%%%%%%%%%%%%%%%%%%%
%%%%%%%%%%%%%%%%%%%%%%%%%%%%%%%%%%%%%%%%%%%%%%%%%%%%%%%%%%%%%%%%%%%
%%%%%%%%%%%%%%%%%%%%%%%%%%%%%%%%%%%%%%%%%%%%%%
Results presented in Figure~\ref{fig4} show that effects of the
electric field on the real part of the effective exciton potential
are relatively small. %On a big scale the differences are so small
%that are almost indistinguishable up to the extremely large
%fields.
It indicates that the corrections to the real part of quasi-bound
exciton energy due to the field must also be small. The result is
consistant with the previous calculations\cite{Nojima88a} for
larger QW, where a trend to a substantial decrease of the Stark
shift was observed with narrowing of quantum well widths up to
$L=50\AA$.

\begin{figure}[tbp]
\includegraphics{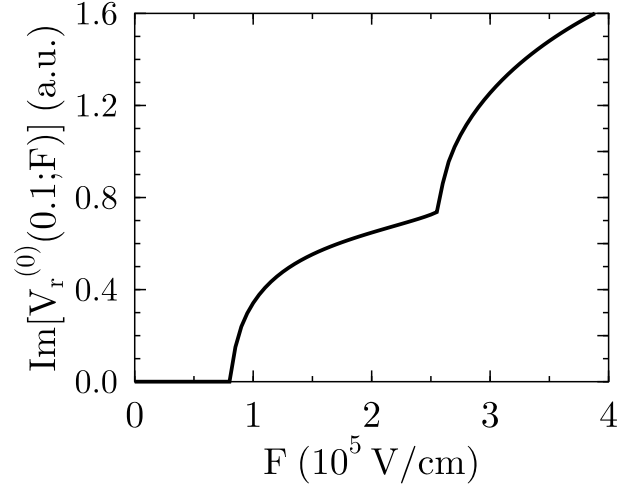}
\caption{The dependence on electric field of an imaginary part of
the effective potential $\overline{V}^{(0)}_r(r;F)$ at fixed value
of radius $r/a_B=0.1$  Data are for $L=20\AA$ AlGaAs/GaAs SQW.}
\label{fig6}
\end{figure}

With increasing applied field the potential changes in the
following way: at small fields (see curves 2,3 on the insert) both
$\beta$s are real and there is no imaginary part at all. As the
field is applied, the electron and the hole are pulled apart by
the field. It weakens the effective attractions and the curve
moves upwards, manifesting a Stark shift for the binding energy
corrections. Curve 3 represents the maximal deflection from the
zero-field potential. It corresponds to the moment of the first
critical field, when the single particle energy for the hole in QW
acquires the width and its tunnelling becomes possible. At this
moment the optical (imaginary) part of the effective potential
also appears. After this the real part of the potential begins
swinging back. The latter can be understood if we have a look at
the real part of $\beta$ in the insert of Fig.~\ref{fig2}. After
the "critical field" is passed, $\beta_R$ also changes its
behavior and even slightly decreases with field growth. The latter
explains the "effective" increasing of the attraction again.
However, since now the connection with the continuum is
established, the physical interpretation is not so obvious. The
imaginary part of the effective potential continues to grow, which
is clearly seen in Fig.~\ref{fig5}. In order to understand the
character of this growth, we fixed the radius of the potential and
investigated the field dependence of the potential strength at
this fixed $r$.  Fig.~\ref{fig6} gives corresponding results for
this dependence. As we can see, it follows the behavior of the
imaginary parts of $\beta_{1,2}$ (compare with Fig.~\ref{fig2}).
The second leap corresponds to the second critical field, when a
new scattering channel, namely, a possibility of electron
tunnelling, is open.
 %%%%%%%%%%%%%%%%%%%%%%%%%%%%%%%%%%%%%%%%%%%%%
 %%%%%%%%%%%%%%%%%%%%%%%%%%%%%%%%%%%%%%%%%%%%%%%%

The smallness of the correction to the real part of the effective
potential suggests that we can try to express this effective
complex potential at finite field through the zero-field
potential, renormalizing some of its parameters.
 To do so, let us look at the asymptotic behavior of the
potential Eq.~(\ref{VeffFr}). One can show that the first two
terms in the asymptotic expansion are
\begin{equation}\label{VeffFr_as}
  \overline{V}^{(0)}_r(r;\kappa_1,\kappa_2,F) \thickapprox
  -1/r+\tilde{d}^2/(2r^3).
\end{equation}
They are the same forms as for zero-field potential [see
Eq.~(\ref{Vr0_asympt})], but with
\begin{equation}\label{dtilda2}
  \tilde{d}^2=\frac{1}{2}\left[
\frac{1}{\kappa_1^2}\frac{1+3\beta_1^2}{(1-\beta_1^2)^2}+
\frac{1}{\kappa_2^2}\frac{1+3\beta_2^2}{(1-\beta_2^2)^2}+
\frac{4\beta_1\beta_2}{\kappa_1\kappa_2(1-\beta_1^2)(1-\beta_2^2)}
  \right]
\end{equation}
It is also easy to see that this new parameter is the average
square of the distance between the electron and the hole in the
$z$ direction, when the electric field applied:
\begin{equation}\label{z2sqf}
\tilde{d}^2=\langle
\chi_1(z_1)\chi_2(z_2)|(z_1-z_2)^2|\chi_1(z_1)\chi_1(z_2)\rangle.
\end{equation}
This parameter is complex now, and its imaginary part provides the
coefficient for the asymptotic of the imaginary part of the
effective exciton potential.  With this finding, it is easy to
understand how we should renormalize the wave numbers
$\kappa_{1,2}$ to adjust corrections in binding energy, induced by
electric field. The easiest way is the following:
\begin{eqnarray}
\frac{1}{\tilde{\kappa}_1}^2&=&\frac{1}{\kappa_1^2}\frac{1+3\beta_1^2}{(1-\beta_1^2)^2}
+\frac{2\beta_1\beta_2}{\kappa_1\kappa_2(1-\beta_1^2)(1-\beta_2^2)},\\
\frac{1}{\tilde{\kappa}_2}^2&=&\frac{1}{\kappa_2^2}\frac{1+3\beta_2^2}{(1-\beta_2^2)^2}+
\frac{2\beta_1\beta_2}{\kappa_1\kappa_2(1-\beta_1^2)(1-\beta_2^2)}.
\end{eqnarray}
%%%%%%%%%%%%%%%%%%%%%%%%%%%%%%%%%%%%%%%%%%%%%%%%%%%%%%%%%%%%%%%%%%%
Then, for small applied fields the potential, Eq. (\ref{VeffFr})
coincides with the modified potential for zero field
Eq.~(\ref{Vr0}) up to the third degree in the electric field
expansion,
\begin{equation}\label{coins}
  \overline{V}^{(0)}_r(r;\kappa_1,\kappa_2,F)\approx
  \overline{V}^{(0)}_r(r;\tilde{\kappa}_1,\tilde{\kappa}_2,0).
\end{equation}
Therefore, the expressions
(\ref{excenergy_sc0},\ref{excenergy_sc1a}) for exciton binding
energy in zero field with renormalized parameter $\tilde{d}^2$ are
valid for non-zero electric fields.
%%%%%%%%%%%%%%%%%%%%%%%%%%%%%%%%%%%%%%%%%%%%%%
\section{Conclusions\label{sec:conclusions}}
We developed a self-consistent complex scaling approach for
calculations of field-induced complex energies of single particle
states as well as of the exciton quasi-energy in QWs. This
technique is a generalization of the self-consistent approach
recently developed by the authors\cite{pon03a} for the exciton in
a QW without the presence of the electric field. For the case of
strong confinement, we represented the trial function in
factorized form and obtained the corresponding coupled
self-consistent equations (\ref{mfeq1}--\ref{mfeq3}), which
include the effective potentials (\ref{efpotr},\ref{efpotz}).
These effective potentials are results of averaging over $z$
coordinates of the Coulomb interaction and the quantum well
potential. We would like to note, that, although the resulting
equations are obtained for a separable wave function, the approach
itself has a broader range of applicability, and can be employed
for different forms of trial functions.

We further applied our approach to the case of a shallow quantum
well. To this goal we developed the model of  QCSE in a
$\delta$-functional quantum well, which allows us to obtain
analytical results for the resonance positions and widths and the
behavior of the modified exciton "ground state" wave function.
This occurs because the single-particle one-dimensional problem in
delta-potential and applied uniform electric field can be solved
exactly. We also compared the exact results for single-particle
complex energies with the  complex scaling variational method
results for a trial function in a simple form of two decaying
exponents with different coefficients of decay. The obtained
energy has both the real and the imaginary parts, and their values
are in a very good agreement with the exact solution for
quasi-energy within a wide range of the parameters. The non-zero
imaginary part gives additional physical insight into the problem.
The simple variational functions for the electron and the hole
allow us to calculate the effective exciton potential. This
potential has real and imaginary (optical) parts. We found that
the real part of the potential is insensitive to changes of
external electric field up to the QCSE ionization threshold. This
allows us to express exciton quasi-bound state at some field
through the renormalized expression for the binding energy in zero
field.
\begin{acknowledgments}
We are grateful to S. Schwarz for reading and commenting on the
manuscript. The work is supported by AFOSR grant F49620-02-1-0305.
\end{acknowledgments}
%\appendix
%%%%%%%%%% Produces the bibliography via BibTeX. %%%%%%%%%%%%
\bibliographystyle{apsrev}
%\bibliography{exciton}

\end{document}